\documentclass[12pt] {iopart}
\usepackage{amsmath}
\usepackage{graphicx}
\begin{document}
\title[Ground state properties and bubble structure]{Ground state properties and bubble structure of the isotopic chains of Z = 125 and 126 using the relativistic mean-field formalism}

\author{Priyanka$^1$, A. Chauhan$^1$, 
M. S. Mehta$^1$, M. Bhuyan$^2$}

\address{$^1$ Department of Physics, Rayat Bahra University, Mohali 140104, India}
\address{$^2$ Center for Theoretical and Computational Physics, Department of Physics, Faculty of Science, Universiti Malaya, Kuala Lumpur 50603, Malaysia}

\ead{priya80593@gmail.com, offlineashish@gmail.com, msmehta@rayatbahrauniversity.edu.in,bunuphy@um.edu.my}

\begin{abstract}
The ground state properties of Z = 125 and 126 nuclei are investigated, taking the isotopic series from the proton to neutron drip-lines. This analysis is conducted using the relativistic mean-field approach with NL3 and the Relativistic-Hartree-Bogoliubov model with DD-ME2 parameterization. The bulk properties under examination include the binding energy, the neutron separation energies, the differential variation of the separation energy, the quadrupole deformation parameter $\beta_2$, and the single-particle energy. We observed the stability at N = 172 and 184 over the isotopic chain for both parameter sets. The quadrupole deformation parameter reveals a shape transition from prolate to spherical and back to prolate with mass number. No signature of a super- and/or hyper-deformed structure is found over the isotopic chain. Furthermore, the analysis is extended to examine the bubble structure, revealing a bubble/semi-bubble structure for a few neutron-rich isotopes.
\end{abstract}


\maketitle

\section{Introduction}
The region of Superheavy masses within the nuclear landscape is a captivating and actively researched domain, both theoretically and experimentally. The objective is to pinpoint the boundaries of nuclear stability and identify potential nuclei that could exist through various combinations of protons and neutrons. With advancements in radioactive ion beam technology, studies in nuclear physics are continually advancing. Recent discoveries, such as nuclei $^{294}$Og (Z = 118) \cite{ogan01} and Z = 117 \cite{ogan10,hofm00}, have provided new avenues for exploring nuclear properties within this mass range. Successful measurements of decay properties for isotopes like $^{293}$Ts and $^{294}$Ts (Z = 117) further strengthen prospects for exploration of the heart of the island of stability within the super-heavy mass region. Laboratory synthesis of elements up to Z = 118 has been achieved thus far through cold fusion reactions (Z = 110-112) \cite{hofm02,zage08,hami13} and hot fusion reactions (Z = 113-118) \cite{ogan06,ogan11,zagr12,ogan13, ogan15}. However, Superheavy nuclei (SHN) represent highly unstable systems with half-lives typically in the order of microseconds or even shorter. Additionally, their very low production yield and cross-section present significant challenges in terms of detection and measurement.

Conversely, theoretical investigations within the Superheavy mass region have been extensive, employing various models, although not all findings are consistent with subsequent magic numbers after Z = 82 and N = 126. The Skyrme-Hartree-Fock approach, using SkP and SLy7 parameter sets, foresees nuclear magicity beyond $Z > 114$, proposing double magicity at Z = 126 and N = 184, and suggests enhanced stability at N = 162 due to deformed shell effects \cite{cwiok96}. In the relativistic Continuum Hartree-Bogoliubov (RCHB) framework, it was suggested that Z = 114 and Z = 120 hold equal likelihood for magic shells in this realm, with magicity varying with isotopes \cite{cwiok99,grei95,rutz97}. Extensive RCHB calculations indicate potential magic numbers for proton (Z = 106, 114, 120, 126, 132, and 138) and neutron (N = 138, 164, 172, 184, 216, 228, 238, and 252) \cite{zhan05}. The Finite-Range Droplet Model (FRDM) identifies the next magic proton number after Z = 82 at Z = 114 and the neutron number at N = 184, while also highlighting significant shell gaps at Z = 104, 106, 108, 110, and N = 162, 164 \cite{lern94}. In the relativistic mean field (RMF) model employing the NL-SH parameter set, double magic numbers of Z = 114 and N = 184 are observed for nuclei below Z = 114 (Z = 100-114) \cite{lala96}. Additionally, see Refs. \cite{patra99,gupt97}, in their comprehensive investigations using RMF, predict the emergence of new magic numbers and the disappearance of some traditional ones, such as relatively larger shell gaps at Z = 92 and 120 compared to Z = 114, and more pronounced shell gaps at N = 138, 184, 258 compared to N = 164, 172, 198 nuclei.

Recently, Bhuyan {\it et al.} \cite{bhuya12}(2012), employing non-relativistic Skyrme-Hartree-Fock (SHF) and relativistic mean field (RMF) models, demonstrated that the magic number for Z = 120, coupled with prominent neutron shell gaps at N = 172, 182/184, 208, and 258. Furthermore, in an in-depth analysis of the ground state properties of nucleus Z = 124, one of the authors of the present study identified magicity at N = 172, 184, and 198 using the relativistic mean-field (RMF) model with the NL3 parameterization \cite{meht15}. Additionally, this investigation predicted another notable feature known as semi-bubble structure in Super Heavy Nuclei (SHN), which was also anticipated in the Skyrme-Hartree-Fock model \cite{bend99}. In prior work of Ref. \cite{patr09} focusing on nucleus Z = 122, stability at N = 172 or 190 (RMF) and N = 182-186 (SHF) was identified using RMF and SHF, a finding recently reaffirmed in studies of Z = 120 with the E-RMF model \cite{jeet23}. Furthermore, Biswal {\it et al.} \cite{bisw14}, using a simple effective interaction, predicted magic nuclei with Z = 114, 120, 126, and N = 182. In the density-dependent relativistic mean-field (RMF) model \cite{saho18}, magic number combinations of Z = 114, 120, 126, and N = 172, 184, 198 were forecasted. The stability of Super-heavy nuclei (SHN) hinges on the interplay among nucleon-nucleon interactions, pairing effects, and neutron-proton configurations. The significance of spin-orbit interaction in determining the single-particle energy shell gap is evident in spectroscopic studies of the nucleus $^{254}$No (Z = 102, N = 152) \cite{herz06}. Hence, it is imperative to look into the internal arrangement of nucleons within Superheavy nuclei. 

The primary objective within the Superheavy region revolves around exploring the center of stability, often referred to as the Island of Stability. Theoretical investigations employing various models yield differing assessments of the additional stability associated with specific proton numbers. Predictions from these models suggest potential magic numbers or centers of stability at Z = 114, 120, and 126, with some sources proposing further magic numbers for neutrons beyond N = 126, possibly at 172 or 184 \cite{lala99,brow90,ring80,meht15}. Beyond magic numbers, other notable ground state properties of nuclei within this mass range encompass quadrupole deformation and/or shape degrees of freedom \cite{zhan22,wang21} and bubble structure, which unveil the unique characteristics of Superheavy nuclei \cite{sing13}. Investigating the impact of higher-order deformation parameters on the ground state of nuclei around $^{270}$ Hs is of particular interest, as it could shed light on how these parameters influence other properties of Superheavy nuclei. As previously stated, Z = 126 stands out as a potential upcoming shell or sub-shell closure and is anticipated to mark the center of the Island of Stability within the Superheavy mass range. Consequently, delving into the neighboring nuclei becomes crucial. Our focus lies in understanding how combinations of protons and neutrons contribute to the relative stability of nuclei in this vicinity. Additionally, we aim to explore the underlying factors that may trigger the formation of bubble or semi-bubble structures within nuclei of this mass range. In the present work we calculate the ground-state properties of the isotopes of nuclei Z = 125 and 126 within the framework of the RMF (NL3) and RHB (DD-ME2) and examine the bubble/semi-bubble structure for a few neutron-rich isotopes. \\
The paper is structured as follows: Sec.\ref{theory} provides an brief explanation of the theoretical framework utilized for computing nuclear bulk properties and discerning bubble structure. The findings are expounded upon in Sec. \ref{result}, while Sec. \ref{summary} encapsulates the current analysis.

\section{Theoretical Formalism}  \label{theory}
In the present calculations, we employ the relativistic mean-field (RMF) model, recognized as a potent tool for effectively describing the ground state properties of nuclei \cite{wale74,sero86,rein86,gamb90,rufa88,hard88,patr97,gupt97,patr97a}. A prominent aspect of the RMF approach is its incorporation of the spin-orbit interaction, arising naturally from the meson-nucleon interaction \cite{horo81}. The model initiates with the Lagrangian density, represented as follows:
\begin{eqnarray}
\label{eqn1}
{\cal L} &=& \bar \psi_i(i\gamma^\mu\partial_\mu - M)\psi_i
+\frac{1}{2}\partial^\mu\sigma\partial_\mu\sigma - \frac{1}{2}m^2_\sigma\sigma^2 - \frac{1}{3}g_2\sigma^3\nonumber\\ 
&-& g_s\bar\psi_i\psi_i\sigma - \frac{1}{4}\Omega^{\mu\nu}\Omega_{\mu\nu} 
+ \frac{1}{2}m^2_\omega V^\mu V_\mu
- g_\omega\bar\psi_i\gamma^\mu\psi_iV_\mu\nonumber\\
&-& \frac{1}{4}\vec B^{\mu\nu}.\vec B_{\mu\nu} 
+ \frac{1}{2}m^2_\rho\vec R^\mu.\vec R_\mu
- g_\rho\bar\psi_i\gamma^\mu\vec\tau\psi_i.\vec R^\mu\nonumber\\
&-&\frac{1}{4}F^{\mu\nu}F_{\mu\nu} 
- e\bar\psi_i\gamma^\mu \frac{(1-\tau_{3})}{2}\psi_i A_\mu. 
\end{eqnarray}
Here, $\sigma$, $V^\mu$, $R^\mu$, and $A^\mu$ represent the fields for $\sigma$, $\omega$, $\rho$, and photons (electromagnetic field), respectively. The $\psi_i$ denote the Dirac spinor of the nucleons, with coupling constants for linear terms, namely $g_\sigma$, $g_\omega$, $g_\rho$, and $\frac{e}{4\pi}=\frac{1}{137}$ for $\sigma$-, $\omega$-, $\rho$-mesons, and photons, respectively. The Greek letter $\vec\tau$ ($\vec\tau_{3}$) signifies the Pauli isospin matrix (the third component of $\tau$) for the nucleon spinor ($\tau_3=-1$ for neutron and +1 for proton). $g_2$ and $g_3$ denote the coupling constants for the non-linear terms of the $\sigma$ meson. M, $m_\sigma$, $m_\omega$, and $m_\rho$ represent the masses of the nucleons, $\sigma$, $\omega$, and $\rho$ mesons, respectively. The field tensors $\Omega^{\mu\nu}$, $R^{\mu\nu}$, and $F^{\mu\nu}$ corresponding to $\omega$-, $\rho$-mesons, and the electromagnetic field, respectively, as appearing in the Lagrangian, are defined as follows:
\begin{eqnarray}
\Omega^{\mu\nu}&=&\partial^\mu\omega^\nu-\partial^\nu\omega^\mu\nonumber\\
R^{\mu\nu}&=&\partial^\mu\vec\rho^\nu-\partial^\nu\vec\rho^\mu
-g_\rho (\vec R^\mu \times \vec R^\nu)\nonumber\\
F^{\mu\nu}&=&\partial^\mu A^\nu-\partial^\nu A^\mu.
\end{eqnarray}
The quantities marked with overhead arrows denote iso-vector properties. The tensor $R^{\mu\nu}$ involves a non-Abelian vector field. However, for simplicity, we approximate $R^{\mu\nu}$ as $\approx \partial^\mu\vec\rho^\nu-\partial^\nu\vec\rho^\mu$. In the RMF model provided in Refs. \cite{niks02,type99}, the meson-nucleon coupling is permitted to have density dependence. This coupling is parameterized using a phenomenological approach \cite{sing14,fuch95,carl00}. The coupling between the nucleon fields and mesons is as follows:
\begin{eqnarray}
g_i(\rho) = g_i\left(\rho_{sat})f_i(x)\right |_{i=\sigma, \omega},
\end{eqnarray}
where,
\begin{equation} 
\label{eqn4}
f_i(x)=a_i \frac{1+b_i\left(x+d_i\right)^2}
{1+c_i\left(x+d_i\right)^2}
\end{equation}
and
\begin{equation}
g_\rho=g_\rho\left(\rho_{sat}\right)e^{a_\rho(x-1)}.
\end{equation}
In this case, there are dependence between the eight real parameters in Eq.(\ref{eqn4}) and the functional $x $=$\rho/\rho{sat}$. The mass of the $\sigma$ meson and coupling parameters, which are independent parameters, were tuned in order to replicate the ground state features of finite nuclei as well as the characteristics of symmetric and asymmetric nuclear matter.
The field equations for mesons and nucleons can be obtained from the relativistic Lagrangian. By enlarging the Boson fields and upper and lower components of Dirac spinors in a deformed harmonic oscillator basis with an initial deformation, these equations are solved. A self-consistent iteration method is used to solve the set of coupled equations numerically.

The standard harmonic oscillator formula is utilized to estimate the center of mass motion, $E_{c.m.}=\frac{3}{4}\left(41A^{-1/3}\right){MeV}$. Describing the nuclear bulk properties of open-shell nuclei necessitates consideration of pairing correlations in both their ground and excited states \cite{kara10}. Several methods have been developed to address pairing effects in the analysis of nuclear properties, including fission barriers \cite{zeng83,nhan12}. Among these methods are the BCS approach, the Bogoliubov transformation, and particle-number conserving methods. The Bogoliubov transformation is widely utilized to account for pairing correlations in the drip-line region \cite{ring96,paar07,lala99,vret99}. For nuclei situated relatively close to the $\beta$-stability line, employing the constant gap BCS pairing approach can yield a reasonably accurate approximation of pairing \cite{doba84}.

The paring correlations will be taken into account in the constant gap approximation taken from the prescription of Madland and Nix \cite{madl88}, 
 \begin{eqnarray}
{\Delta}_{p}= r b_sZ^{-1/3}e^{(sI-tI^2)} \nonumber\\
{\Delta}_{n}= r b_sN^{-1/3}e^{-(sI+tI^2)}
\end{eqnarray}
Here, we have $r=5.72$ MeV, $s=0.118$, $t=8.12$, $b_s=1$, and $I=(N-Z)/(N+Z)$. The proton ($\Delta_p$) and neutron ($\Delta_n$) gap parameters are zero for closed shell nuclei. However, for non-magic N or Z, $\Delta_p$ and $\Delta_n$, necessary for obtaining occupation probabilities in the expression for densities, are derived here following the prescription of Madland and Nix \cite{madl88}. Other methods of treating the pairing interaction will also be considered based on the suitability of the mass region in the nuclear chart.

Moreover, the relativistic Hartree-Bogoliubov (RHB) model \cite{niks14} stands as a robust theoretical framework extensively utilized in nuclear physics for elucidating the properties and behavior of atomic nuclei. This model merges the principles of the Hartree-Fock approach and the Bogoliubov transformation within a relativistic context, capturing the effects of both mean fields and particle-particle correlations. Within the RHB model, nucleons are treated as quasi particles interacting through effective mean fields, derived from a relativistic Lagrangian. These mean fields, including scalar and vector potentials, are self-consistently determined by solving the coupled set of relativistic mean field equations. The RHB model incorporates the Dirac equation for nucleons to account for their relativistic dynamics, facilitating the description of both bound and unbound states of nuclei. The field equations under the Hartree approximation for self-consistent mean field are expressed as follows:
\begin{eqnarray}
\begin{bmatrix}
        \hat h_D-m-\lambda & \hat\Delta\\
        -\hat\Delta^*& -\hat h_D + m + \lambda\\
\end{bmatrix}
\begin{bmatrix}
        U_k(r)\\
        V_k(r)\\
\end{bmatrix} = E_k
\begin{bmatrix}
        U_k(r)\\
        V_k(r)\\
\end{bmatrix}.
\end{eqnarray}
In this context, $\hat h_D$ represents the single-nucleon Dirac Hamiltonian, with $m$ denoting the nucleon mass. The functions $U_k(r)$ and $V_k(r)$ are localized functions of $r$, where $\lambda$ stands for the chemical potential, positioned below the continuum limits if the pairing field $\hat\Delta$ is both diagonal and constant.

\section{Results and Discussion} \label{result}
In this study, we explore the ground state properties of Super Heavy Nuclei (SHN) with odd and even mass isotopes of Z = 125 and 126, using the relativistic mean field (RMF) model with NL3 parameter set and the Relativistic Hartree-Bogoliubov (RHB) approach. We then contrast our findings with recently predicted values from the Macroscopic-microscopic finite range droplet model (FRDM)\cite{jach21} and the nuclear mass table employing the global mass formula (WS4) \cite{ning14}. For fermions and bosons, we set the oscillator shell numbers as $N_F = 12$ and $N_B = 20$, respectively. The comparative analysis of our outcomes is delineated in subsequent sections.
\begin{figure}[h]
\centering
\includegraphics[width=0.9\columnwidth,clip=true]{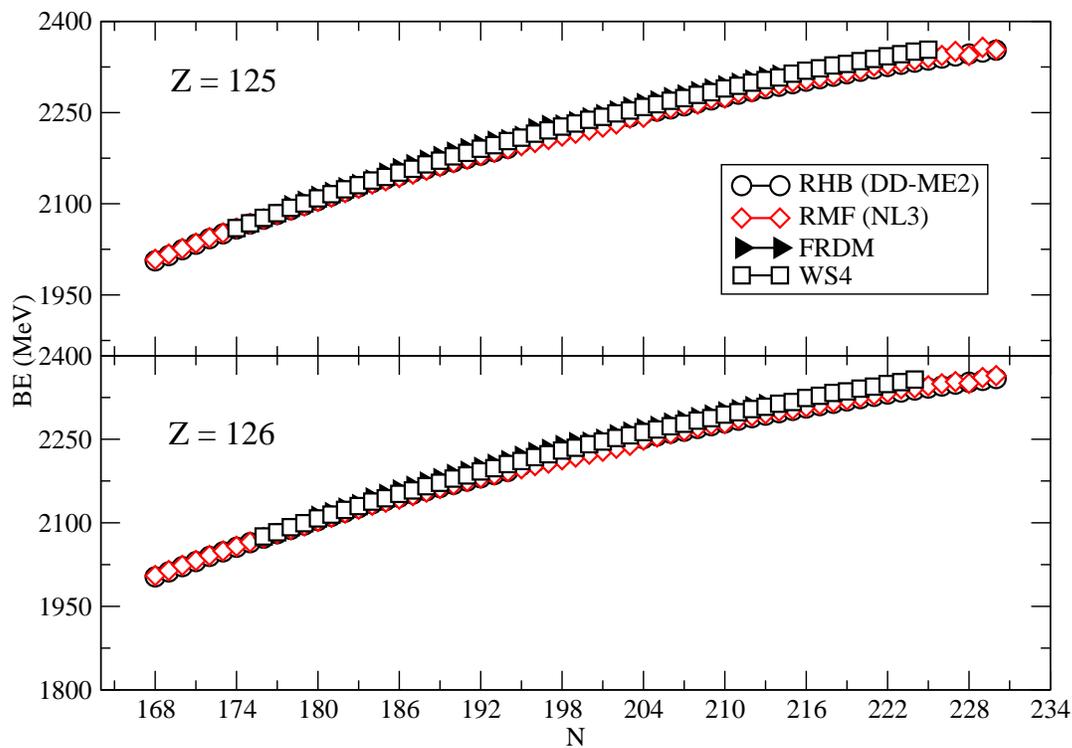}
\caption{\label{fig1} The Binding Energy of the isotopic series of Z = 125 and 126 nuclei from RMF(NL3) and RHB(DD-ME2) are compared with FRDM \cite{moll16} and WS4 \cite{ning14}.}
\end{figure}
\subsection{Binding Energy}
Using the relativistic mean field model with NL3 parameterization and the relativistic Hartree-Bogoliubov model with DD-ME2 parameterization, we calculate the binding energy of the isotopic chain of Z=125 and Z=126 nuclei. We compare nuclear binding energies across the isotopic series from the proton drip-line to the neutron drip-line (N = 168-220) for nuclei with Z=125 and Z=126 with the predictions from FRDM  \cite{moll16} and WS4\cite{ning14}. From Fig. \ref{fig1}, we observe minor discrepancies over the heavier mass region. Conversely, our models exhibit strong agreement with FRDM and WS4 in the lighter mass region. For instance, considering a spherical ($\beta_2$ = 0.001) $^{309}$125 nucleus, the binding energy is 2138.34 MeV in FRDM and 2138.1880 MeV in WS4, while it is 2132.575 MeV in RMF and 2134.965 MeV in RHB. Similar patterns are observed for the binding energies of the $^{323}$125 isotope, with values in RMF (2211.4 MeV) and RHB (2217.5 MeV) showing close agreement, and corresponding figures in FRDM (2228.71 MeV) and WS4 (2226.9548 MeV) exhibiting comparable results. No super-deformed isotope is predicted by our current computations. Fig. \ref{fig1} illustrates the slightly larger variations in binding energy at the heavier side compared to the lighter side. To validate the consistency of our findings, we compute the binding energy for the isotopic chain of Z = 126, as shown in Figure 1. For each isotope in the series, both models predict binding energies that closely align with each other.
\begin{figure}
\centering
\includegraphics[width=0.9\columnwidth,clip=true]{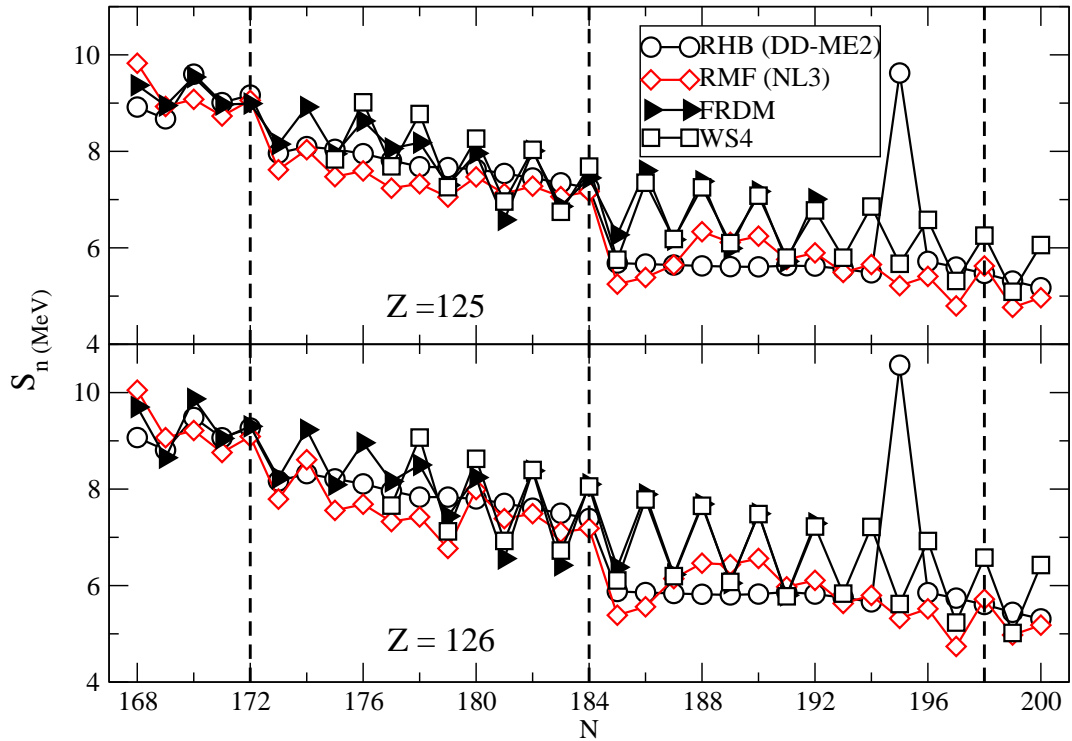}
\caption{\label{fig2} The One neutron separation energy as a function of Neutron number from RMF and RHB theory with NL3 and DD-ME2 parameter for the isotopic series of Z = 125 and 126 nuclei are compared with FRDM \cite{jach21} and WS4 \cite{ning14}.}
\end{figure}
\begin{figure}[h]
\centering
\includegraphics[width=0.9\columnwidth,clip=true]{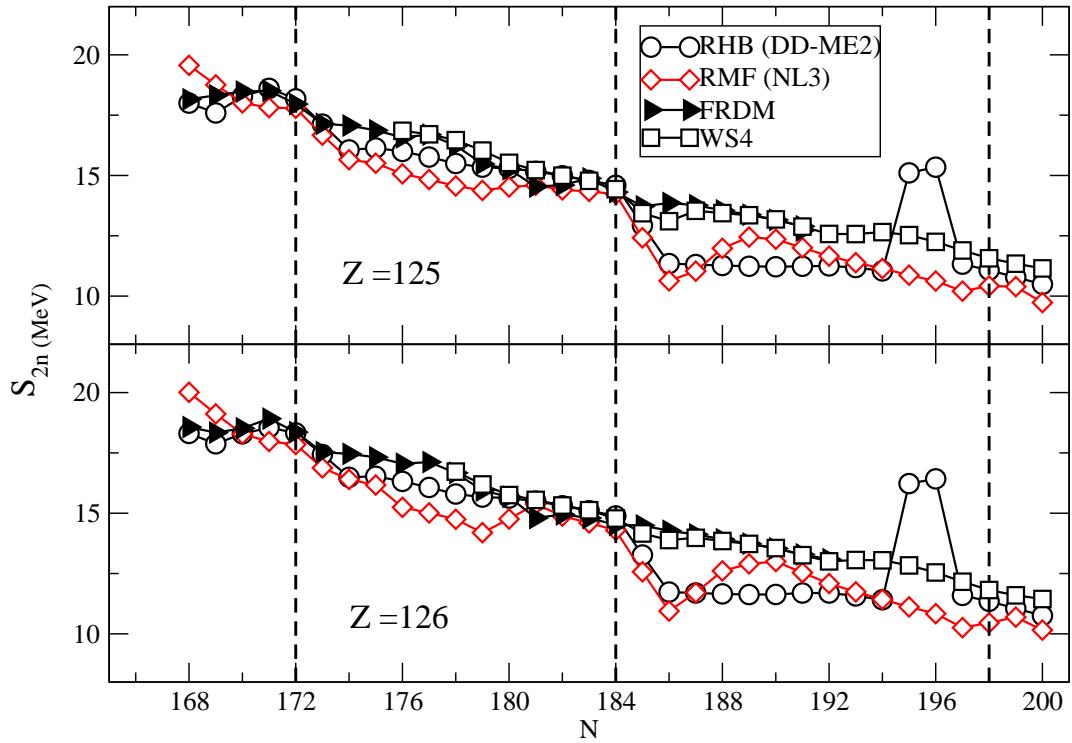}
\caption{\label{fig3} Same as fig.2 but for two neutron separation energy of Z = 125 and 126 nuclei are compared with FRDM \cite{jach21} and WS4 \cite{ning14}.}
\end{figure}

\subsection{Separation Energy}
The separation energies for one and two neutrons are calculated by the difference of binding energy of two isotopes using relations:
\begin{eqnarray}
S_{2n}(N,Z) = BE(N,Z) - BE(N-2,Z)\nonumber\\  
	 S_{n}(N,Z) = BE(N,Z) - BE(N-1,Z)\nonumber
\end{eqnarray}
Figs. \ref{fig2} and \ref{fig3} depict the results of one and two neutron separation energies for the isotopic chain of Z = 125 and 126 nuclei.  The results obtained from RMF and are in good agreement with those from FRDM \cite{jach21}, and WS4 \cite{ning14}. The findings reveal a notable decrease in energy at N = 172 and N = 184 for both one and two neutron separation energies, signifying enhanced stability of nuclei at these neutron numbers. Specifically, the decline in RMF for S$_n$ at N = 172 and 184 are 1.438 MeV and 1.92 MeV, respectively, while for S$_{2n}$ it is 1.114 MeV and 1.81 MeV. In the RHB model, the reduction in S${n}$ energy at N = 172 and N = 184 is 1.204 MeV and 1.56 MeV, respectively, and for S${2n}$ it is 1.048 MeV and 1.664 MeV. For Z = 126, the decrease in RMF for S${n}$ at N = 172 and N = 184 is 1.297 MeV and 1.80 MeV, respectively, while for S${2n}$ it is 0.96 MeV and 1.72 MeV. In the RHB model, the decline in S${n}$ energy for N = 172 and N = 184 is 1.098 MeV and 1.5126 MeV, respectively, and for S${2n}$ it is 0.8969 MeV and 1.6218 MeV. The calculated results are in good agreement with those from FRDM and WS4. Notably, FRDM does not exhibit a sudden decline in S$_{2n}$ energy at N=184, and the calculated values indicate a separation energy of approximately 2 MeV.
\begin{figure}
\centering
\includegraphics[width=0.9\columnwidth,clip=true]{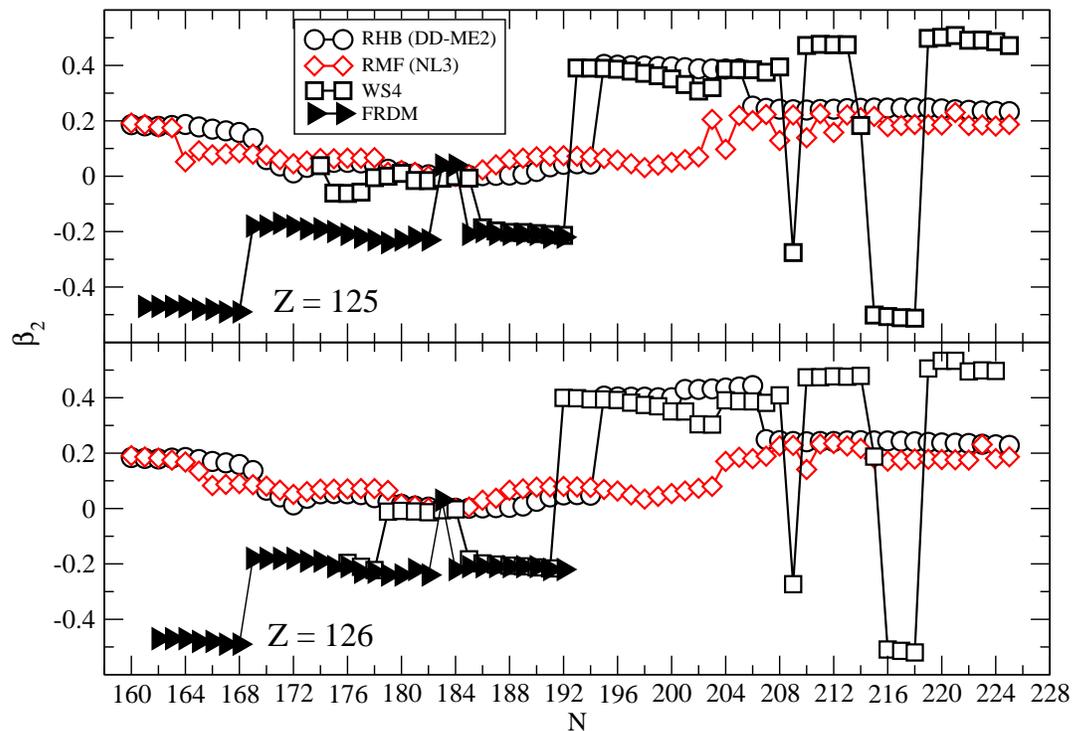}
\caption{\label{fig4} The values of quadrupole deformation parameter, $\beta_2$ for isotopic series of Z = 125 and 126 nuclei are compared with FRDM and WS4. The black filled circles represent the FRDM data \cite{jach21} and square represent the WS4 data \cite{ning14}.}
\end{figure}

\subsection{Quadrupole Deformation Parameter}
The configuration of a nucleus is dictated by the quadrupole deformation parameter $\beta_2$, where a positive (negative) value of $\beta_2$ signifies a prolate (oblate) shape, while $\beta_2 = 0$ indicates a spherical nucleus. The parameter $\beta_2$ is determined through the following relations:
\begin{eqnarray}
Q=Q_n+Q_p=\sqrt\frac{16\pi}{5}\frac{3}{4\pi}AR_0^2\beta_2, \nonumber
\end{eqnarray}
with R$_0$=1.2A$^\frac{1}{3}$(fm). Fig. \ref{fig4} illustrates the results of the quadrupole deformation parameter for isotopes with Z = 125 nuclei, indicating that the majority of isotopes are either spherical or nearly spherical. In RMF calculations, isotopes become prolate beyond N = 200, while in RHB calculations, they exhibit prolate deformation beyond N = 194. However, in the current computations, no isotope displays significant deformation (hyper-deformed shape). Recent simulations using FRDM suggest that isotopes ranging from $^{285}$ to $^{293}$125 are super deformed (oblate) \cite{jach21}, while those from $^{294}$ to $^{307}$125 and from $^{310}$ to $^{317}$125 are moderately deformed. In WS4, the isotopes from $^{298}$ to $^{310}$125 and $^{303}$ to $^{310}$126 are nearly spherical, whereas those from $^{317}$ to $^{333}$125 and $^{317}$ to $^{334}$126 are prolate deformed. Nuclei $^{308}$ and $^{309}$125 are spherical or near-spherical, consistent with the models employed here. The structure of isotopes of nucleus Z = 126 is predicted to be the same in both models used in the present calculations. The magnitude of the quadrupole deformation parameter, $\beta_2$, ranges between 0.2 to 0.5 (oblate deformation) in available FRDM data published very recently \cite{jach21}, and from 0.03 to 0.3 in WS4 \cite{ning14}. The nuclei from $^{310}$ to $^{318}$125 and $^{311}$ to $^{319}$126 are predicted to be near spherical or spherical in both models used here. However, there is a difference in the structure of these isotopes ($184 < N < 192$) in FRDM between two different versions \cite{moll16,jach21} for the isotopes of nuclei Z = 125 and 126. The complex structure of nuclei in this mass region requires further detailed investigation to provide deeper insights into shape evolution.

\begin{figure}[b]
\centering
\includegraphics[width=16.0cm,height=11.0cm,clip=true]{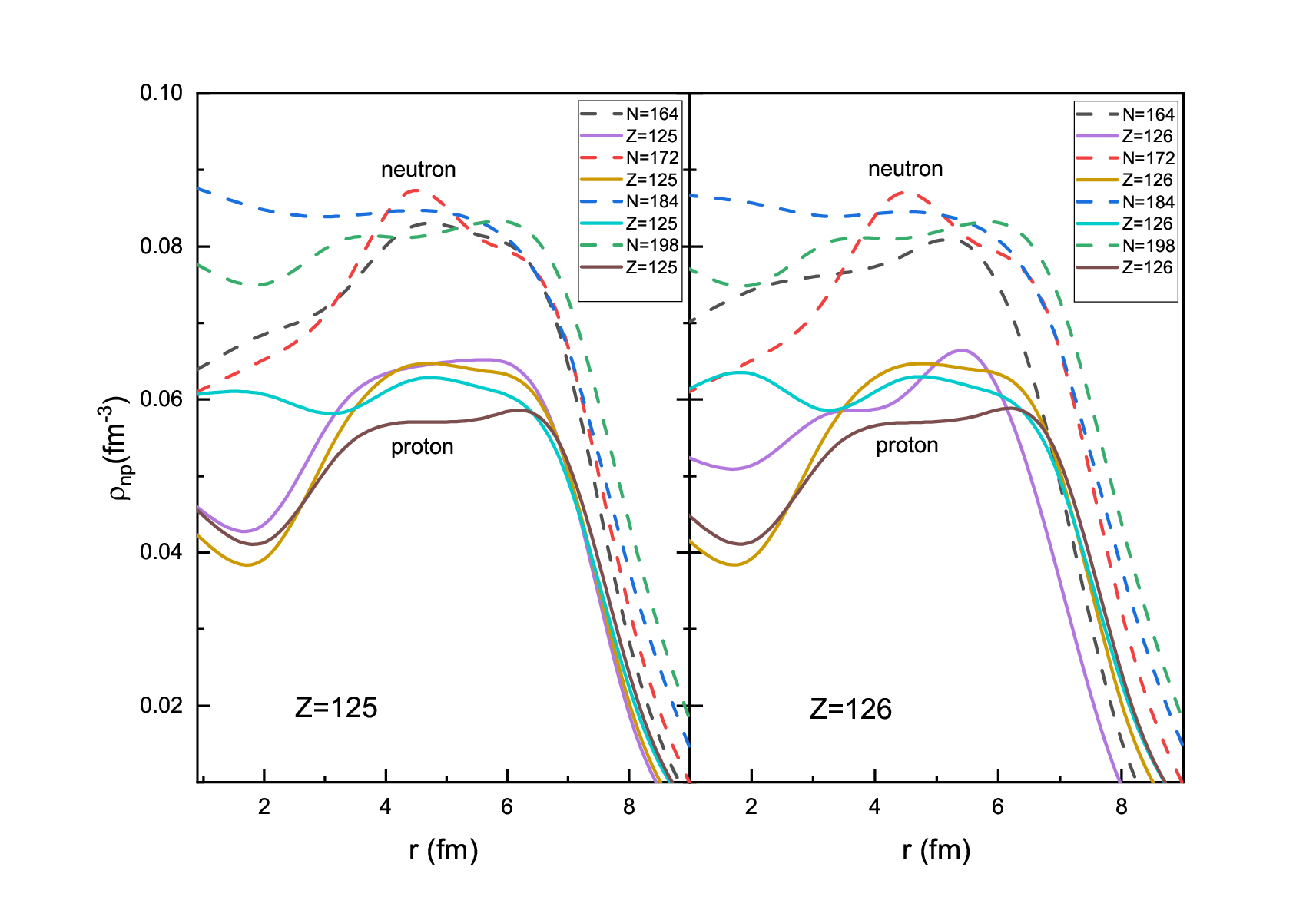}
\caption{\label{fig5} The RMF (NL3) density profile of the isotopes of Z = 125 and 126 nuclei.}
\end{figure}
\subsection{Density Profile and Bubble Structure}
The density profile of a nucleus offers crucial insights into its internal structure, revealing the distribution of nucleons within. Typically, nucleon density peaks at the center of the nucleus, but some nuclei exhibit lower density at or near the center, leading to intriguing structural characteristics. In our current study, we compute neutron and proton densities for selected isotopes of Z = 125 and 126 nuclei using RMF (NL3), as depicted in Figure \ref{fig5}. From the figure, it becomes apparent that the density of Z = 125 isotopes at N = 164, 172, and 198 is lower near the center, while it remains nearly constant in the Z = 125, N = 184 nucleus. Additionally, the structure of Z = 126 isotopes is predicted to be nearly identical, as evidenced by the density profile shown in Figure \ref{fig5}.

\begin{figure}[t]
\centering
\includegraphics[width=15.0cm,height=12.0cm,clip=true]{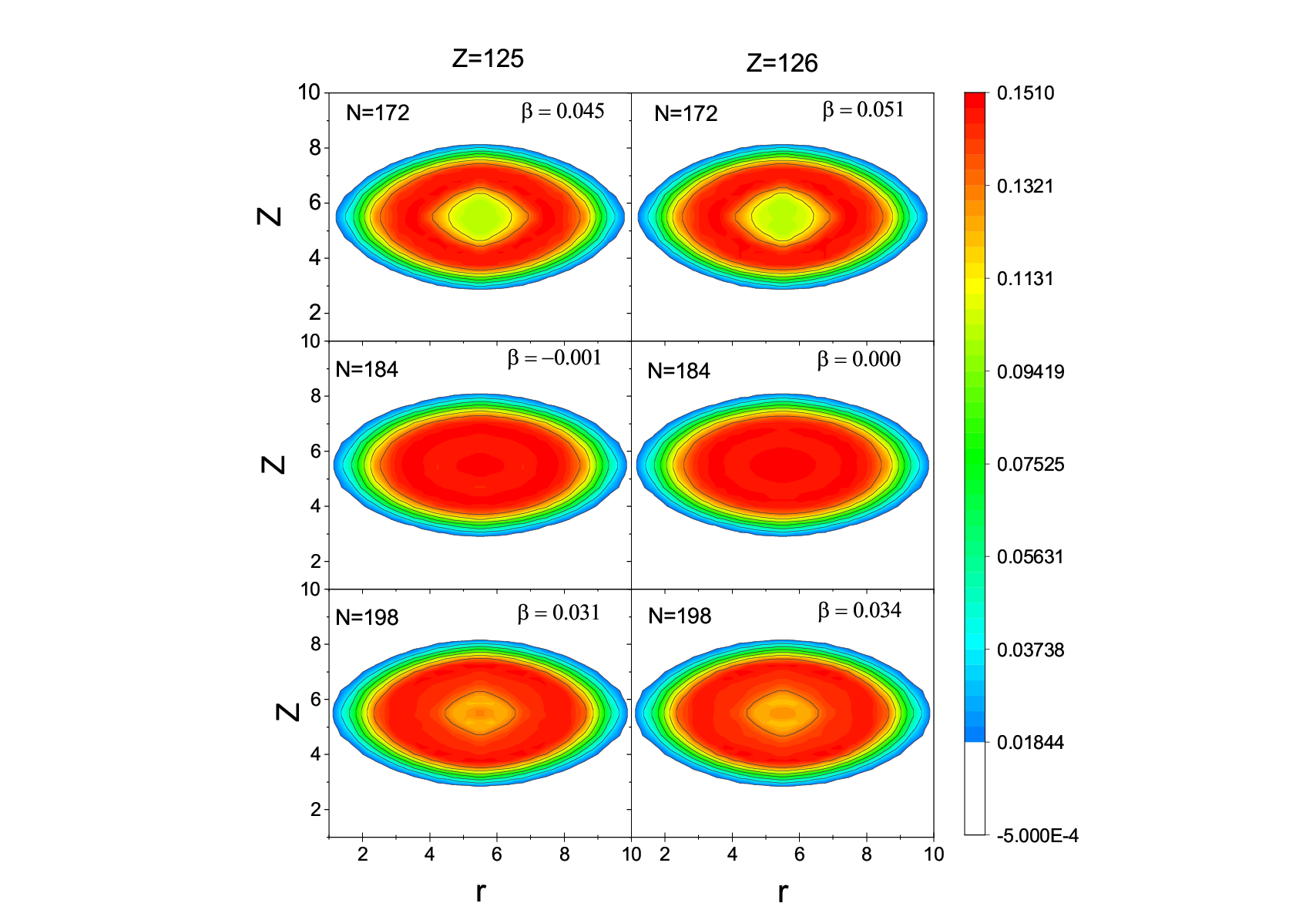}
\caption{Density contours  of the ground state density distribution for the isotopes of Z = 125 and 126 (N = 172, 184 and 198) nuclei with RMF (NL3) model.}
\label{fig6}
\end{figure}

\begin{figure}
\centering
\includegraphics[width=9.0cm,height=11.0cm,clip=true]{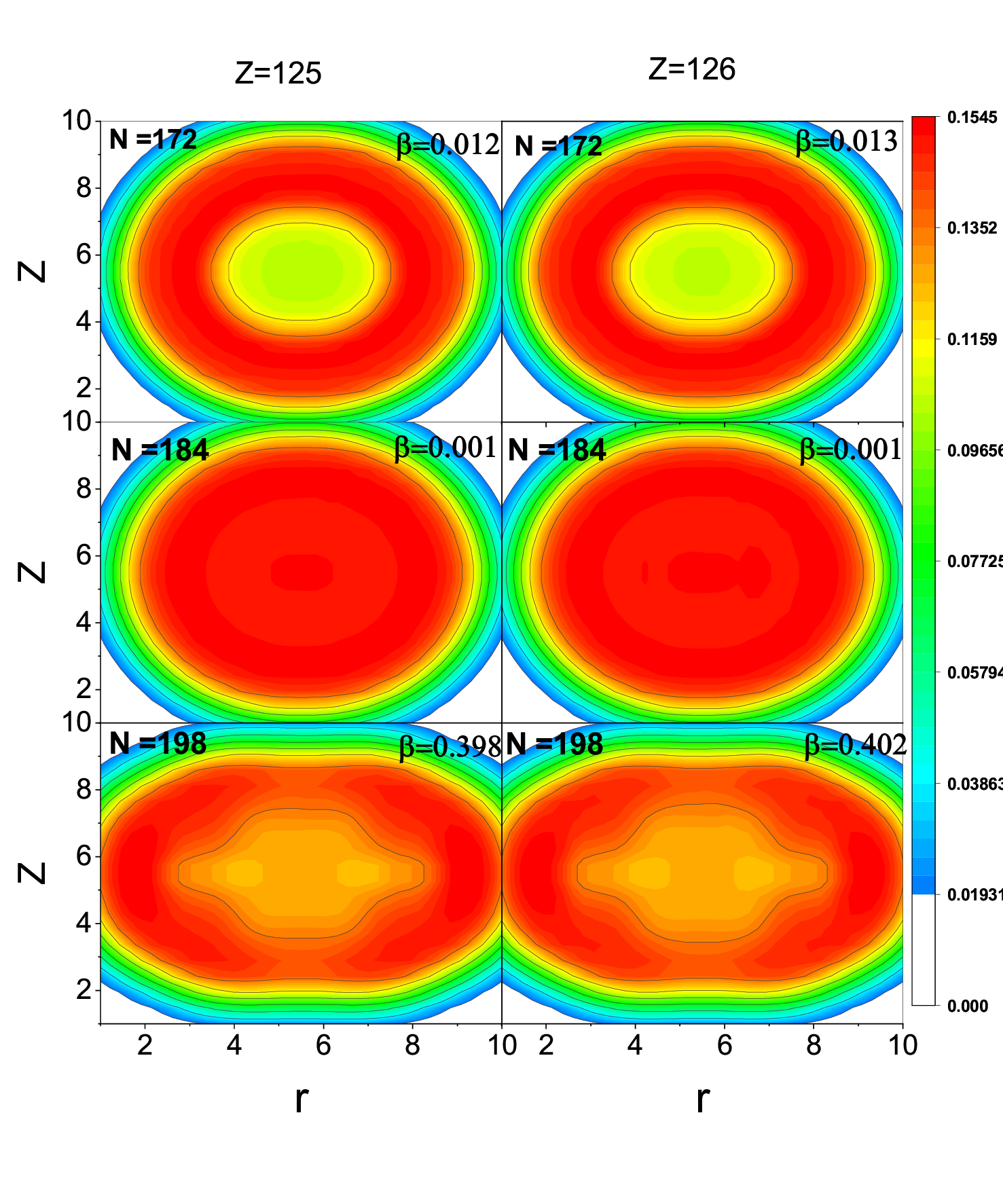}
\caption{Density contours  of the ground state density distribution for the isotopes of Z = 125 and 126 (N = 172, 184 and 198) nuclei with RHB (DD-ME2) model. }
\label{fig7}
\end{figure}

To provide further insight into the structure of these isotopes, we generate contour plots for selected nuclei Z = 125, N = 172, 184, and 198, as displayed in Figures \ref{fig6} and \ref{fig7}. For instance, isotopes such as $^{297}$125, $^{323}$125, $^{298}$126, and $^{324}$126 with N = 172 and 198 for Z = 125 and 126 nuclei exhibit low density at the center, with both models indicating a shell-like structure. Conversely, nuclei like $^{309}$125 and $^{324}$126 demonstrate nearly constant or uniform density up to 6.0 fm, as evident from Figs. \ref{fig6} and \ref{fig7}. Moreover, nuclei $^{297}$125 and $^{298}$126 are predicted to possess a bubble structure, while $^{323}$125 and $^{324}$126 are anticipated to exhibit a semi-bubble type structure. Previously, the extensively studied Superheavy nucleus $^{292}$120 was projected to have a semi-bubble structure \cite{dech99, pei05}, and recently, nuclei such as $^{288,294,296,322}$124 were also suggested to possess a bubble-type structure in a study conducted by one of the present authors \cite{meht15}.

To shed light on the decrease in density magnitude at the central region of these nuclei, we compute the depletion factor (DF) using the relation \cite{saho20, chu10} as follows:
\begin{eqnarray}
DF = \frac{\rho_{\mbox{max.}}-\rho_{\mbox{center}}}{\rho_{\mbox{max.}}} \nonumber
\end{eqnarray}
In this equation, $\rho_{\text{max}}$ and $\rho_{\text{center}}$ represent the maximum and central densities, respectively. The depletion factor is determined to be 30.03\% and 34.12\% for neutrons and protons, respectively, in the $^{297}$125 nucleus, and 30.26\% and 34.29\% for neutrons and protons, respectively, in the $^{298}$126 nucleus using the RMF (NL3) model. Meanwhile, in the RHB (DD-ME2) model, the depletion factor for these nuclei, i.e., $^{297}$125 and $^{298}$126, is 33.0\% and 32.6\%, respectively.

Similarly, for nuclei $^{323}$125 and $^{324}$126, the depletion factor is approximately 12.0\% for both neutrons and protons in the RMF (NL3) model, whereas it is 17.2\% in the RHB (DD-ME2) model. In nuclei $^{309}$125 and $^{310}$126, there is no decrease in density at the central region in both the RMF (NL3) and RHB (DD-ME2) models. The contour plots illustrating the decrease in density at the central region of these selected nuclei are depicted in Figures \ref{fig6} and \ref{fig7}, clearly indicating a bubble or semi-bubble type structure. The $\beta_2$ values for these isotopes with N = 172, 184, and 198 are approximately 0.04, 0.00, and 0.03, respectively, indicating that these nuclei are either spherical or nearly spherical.

Recently, another relation for central depression is used by \cite{schu17} which is given by, 
\begin{eqnarray}
\bar\rho_{\mbox{t,c}} = \frac{\rho_{\mbox{t,av}}-\rho_{\mbox{t,c}}}
{\rho_{\mbox{t,av}}}. \nonumber
\end{eqnarray}
In this equation, $t = (\pi, \nu)$, and $\rho_{t, c}$ represents the central density, while $\rho_{t, \text{av}} = \frac{N_t}{\frac{4}{3}\pi R_{\text{rms}}^3}$ denotes the average density of the nucleus. Here, we employ the root mean square (rms) radius instead of the diffraction radius $R_d$. It is believed that the rms radius remains unaffected by the diffraction radius \cite{frie82}. The depletion factor is determined to be 21.38\% and 23.24\% for neutrons and protons, respectively, in the $^{297}$125 nucleus, and in the $^{298}$126 nucleus, it amounts to 21.40\% and 23.23\% for neutrons and protons, respectively, in the RMF (NL3) model. Furthermore, for the $^{309}$125 and $^{310}$126 nuclei, it is 19.09\% for neutrons and 21.78\% for protons, and 19.22\% for neutrons and 21.79\% for protons, respectively. Similarly, for the nuclei $^{323}$125 and $^{324}$126, the depletion factor is approximately 19.06\% for neutrons and 22.07\% for protons.

The larger depletion factor observed for protons compared to neutrons suggests that the Coulomb force alone cannot account for the reduced density. Both formulas are affected by single-particle effects, with the decreased density attributed to the behavior of individual particles. The role of the Coulomb force is therefore considered secondary in the formation of bubble structures. Upon analyzing various aspects of bubble structure, it becomes evident that the formation of such structures in Superheavy nuclei is primarily influenced by particle distribution. The Coulomb force cannot be solely responsible for the decreased density observed in nuclei such as $^{297}$125 and $^{298}$126, as other nuclei with the same number of protons do not exhibit a decrease in density in the central region.

\begin{figure}
\centering
\includegraphics[width=13.0cm,height=9.0cm,clip=true]{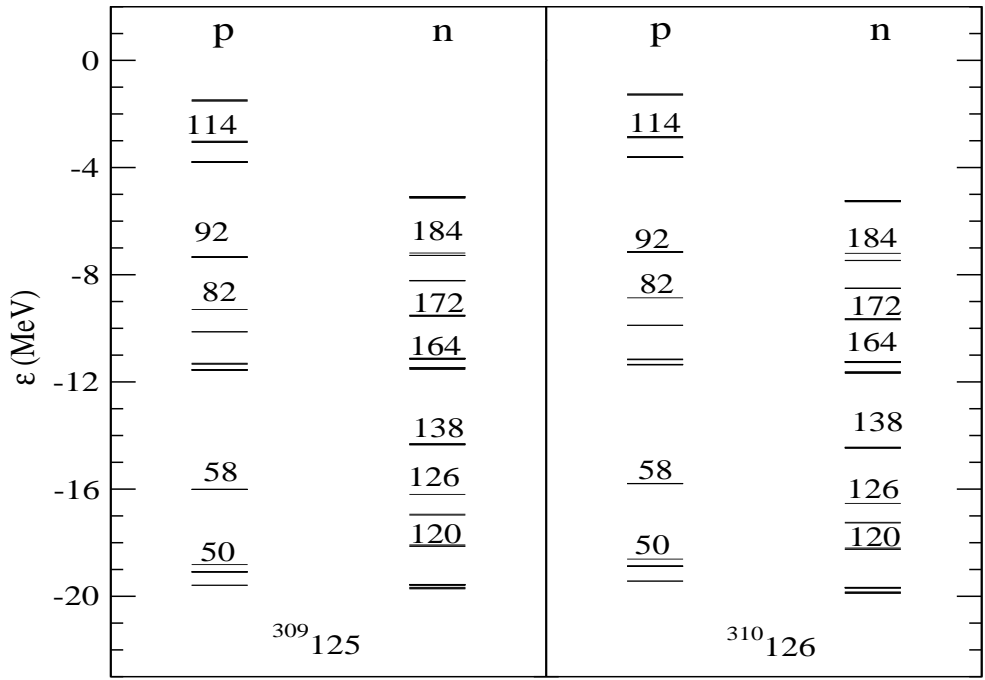}
\caption{The Single particle energy  levels for $^{309}$125 (N = 184) and $^{310}$126 (N = 184) nuclei using RMF (NL3) model. } 
\label{fig8}
\end{figure}
\begin{figure}
\centering
\includegraphics[width=13.0cm,height=9.0cm,clip=true]{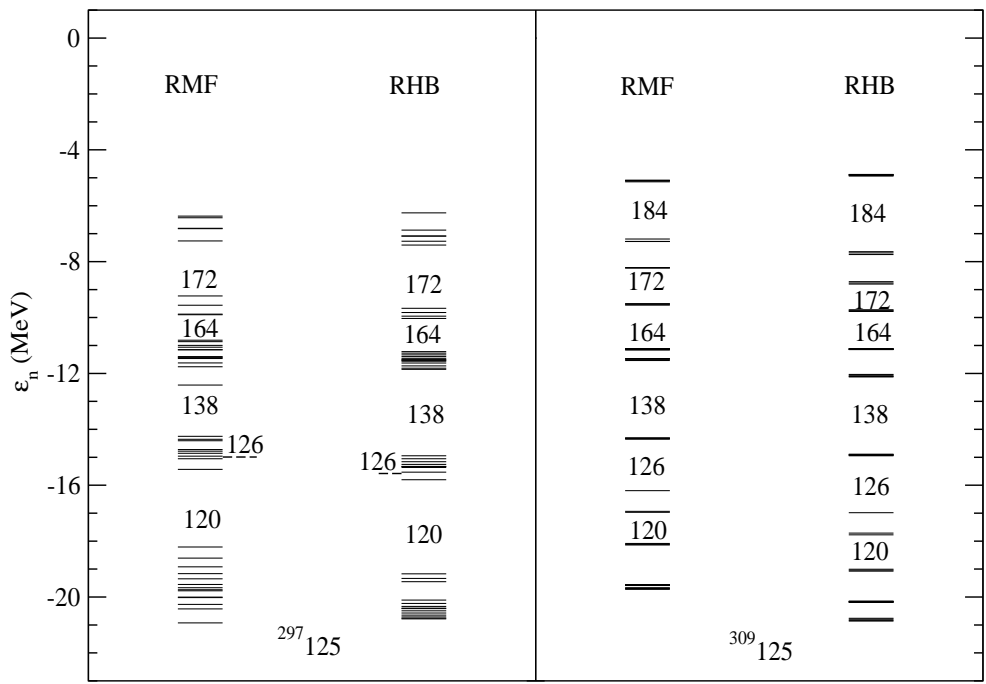}
\caption{The Single particle energy  levels for $^{297}$125 (N = 172) and $^{309}$125 (N = 184) nuclei using RMF (NL3) and RHB (DD-ME2) model.}
\label{fig9}
\end{figure}
\subsection{Single Particle Energy}
To verify the shell/sub-shell closures, we plot the single-particle energy levels for $^{309}$125 and $^{310}$126 nuclei, as depicted in Fig. \ref{fig8}. A slight ($\sim 1.5$ MeV) shell gap is observed at Z = 114, while the gaps at N = 126 and 184 are approximately $2.5$ MeV. However, the shell gap at N = 120 remains relatively small ($\sim 1.0$ MeV). The stability of N = 164 appears to slightly dominate over N = 172. Similar energy levels are observed for $^{297}$125 (N = 172) and $^{309}$125 (N = 184), as illustrated in Fig. \ref{fig9}. In nucleus $^{297}$125, neutron shell gaps are evident at N = 120, 138, 164, and 172, while in nucleus $^{309}$125, shell gaps appear at N = 120, 126, 138, 164, 172, and 184. Notably, the shell gap at N = 126 is not visible at all in nucleus $^{297}$125, as predicted by both models employed here. The sequence of shell gaps in isotopes $^{298}$126 (N = 172, Z = 126) and $^{310}$126 (N = 184, Z = 126) corresponds similarly. This sequence of shell gaps for isotopes of nuclei Z = 125 and 126 aligns with the findings in reference \cite{cwiok96}.

\section{Conclusion} \label{summary}
In this study, we utilized the relativistic Hartree Bogoliubov (RHB) and relativistic mean field (RMF) models parameterized with NL3 and DD-ME2, respectively, to analyze the bulk properties of isotopes belonging to Z = 125 and 126. A comprehensive comparison was conducted with results obtained from the macroscopic-microscopic finite range droplet model and the global mass formula WS4. Our findings regarding binding energy, separation energy, and density distribution exhibited strong agreement with those from FRDM and WS4. In terms of quadrupole deformation, our computations in the lighter mass range of the series revealed a transition from prolate to spherical and back to prolate with increasing mass number. Conversely, FRDM suggested the presence of superdeformed nuclei. Our analysis indicated that most isotopes predominantly exhibited a spherical shape, with deviations observed at higher neutron levels. While the RMF and RHB models exhibited similar binding energies, slight differences were noted on the neutron-rich side when compared to FRDM and WS4. A significant drop in separation energy was observed at N = 172 and N = 184, indicative of the presence of visible shell gaps, although their magnitude was relatively small, approximately 1.5 to 2.0 MeV. Notably, nuclei such as $^{297}$125 and $^{298}$126 were projected to possess a bubble or semi-bubble structure. Ongoing detailed investigations into the potential decrease in density at the center of these nuclei promise intriguing insights.
\section*{References}

\label{bibby}

\end{document}